# Guidance of the resonance energy flow in the mechanism of coupled magnetic pendulums

Valery N. Pilipchuk[a], Krystian Polczyński[b], Maksymilian Bednarek[b], and Jan Awrejcewicz[b*]

[a]Department of Mechanical Engineering
Wayne State University
5050 Anthony Wayne Drive, Detroit, Michigan 48202, U.S.A.
pilipchuk@wayne.edu [https://orcid.org/0000-0001-7051-3539]

[b]Department of Automation, Biomechanics and Mechatronics
Lodz University of Technology
90-924 Lodz, 1/15 Stefanowskiego Str., A22, Poland
krystian.polczynski@dokt.p.lodz.pl [https://orcid.org/0000-0002-1177-6109]
maksymilian.bednarek@dokt.p.lodz.pl [https://orcid.org/0000-0002-7669-119X]
*corresponding author: jan.awrejcewicz@p.lodz.pl [https://orcid.org/0000-0003-0387-921X]

**Abstract**

This paper presents a methodology of controlling the resonance energy exchange in mechanical system consisting of two weakly coupled magnetic pendulums interacting with the magnetic field generated by coils placed underneath. It is shown that properly guided magnetic fields can effectively change mechanical potentials in a way that the energy flow between the oscillators takes the desired direction. Studies were considered by using a specific set of descriptive functions characterizing the total excitation level, its distribution between the pendulums, and the phase shift. The developed control strategies are based on the observation that, in the case of antiphase oscillation, the energy is moving from the pendulum subjected to the repelling magnetic field, to the oscillator under the attracting field. In contrast, during the inphase oscillations, the energy flow is reversed. Therefore, closed-loop controller requires only the information about phase shift, which is easily estimated from dynamic state signals through the coherency index. Advantage of suggested control strategy is that the temporal rate of inputs is dictated by the speed of beating, which is relatively slow compared to the carrying oscillations.

# 1. Introduction

Each of us has encountered an object called a pendulum at least once in a lifetime. Pendulums have been accompanying people for centuries, e.g., in the form of well-known a children's swing. However, they also played more advanced and functional roles. In antiquity (132 AD), a pendulum mechanism was developed in China to alert people to an impending earthquake [1,2]. Nevertheless, the pendulum has gained the greatest popularity as a component of the self-regulating mechanism, commonly known as an escapement, used in clocks. The first design of a modern pendulum clock was developed by Huygens [3,4] in 1673.

Due to the rapid development of technology and the desire to design synergistic systems by connecting various fields of technology to broaden their applications, more and more purely mechanical systems are subjected to the electromagnetic field. The example of such a synergistic system that revolutionized the industry is an electric motor [5,6]. Taking into account the construction of the electric stepper motor, we can find an analogy to pendulum series subjected to magnetic forces. It can be imagined that a rotor is made of many pendulums with magnet fixed at the end of the arms and they are subjected to the changing magnetic field generated by the stator coils [7,8]. Research on the dynamics of the so-called *magnetic pendulums* can shed light on a lot of valuable information on mechatronics devices, such as mentioned electric motors, but also on the magnetic levitation systems used in the MAGLEV railway, magnetic bearings [9], or the mechanisms applied in energy harvesters [10].

In the present work, we suggest to guide the direction of energy flow actively by generating appropriate magnetic fields around the interacting mechanical oscillators, that is, two magnetic pendulums weakly coupled with a torsion spring. Although such an approach extends the class of systems from mechanical to electromechanical, there are certain advantages. First, the corresponding design can be done without mechanical modifications incorporating passive components (i.e magnets and coils) into the entire structure. It is shown that the magnetic field affects the main system parametrically by changing its spectral properties in a proper way. Second, electromagnetic elements are relatively light and more flexible for different adjustments. Finally, it will be shown that the magnetic field can serve as a reliable energy flow guiding tool regardless initial states of the mechanical structure.

Mechanisms of energy propagation, trapping, and dissipation represents one of the most fundamental physical problems on both micro- and macro-levels. In particular, such problems may occur when designing molecular structures with desired *targeted energy transfer* (TET) properties [11–13], vibrational energy harvesters [14,15], or mechanical energy absorbers for controlling the structural dynamics [16,17]. In the latter case, the intent is to create a relatively light device for either harvesting or just absorbing and dissipating the energy from the main structure in an irreversible way. The necessity to guide the energy flow is typically due to design constraints since absorbing devices cannot be usually installed over the entire structure. Note that physical principles of energy absorption and vibration suppression are different. Namely, the vibration suppression typically deals with stationary or at least quasi stationary process of forced vibrations with the intent to achieve appropriate spatial shapes of the amplitude-frequency response. In contrast, the energy absorption is a nonstationary process usually triggered by initial perturbations. Such a specific appears to create some major problems faced by the design of passive energy absorbers. First, the intent to intensify the energy flow between the donor and recipient based on the internal resonance effect may work in either direction in such a way that the recipient may become a donor. Since both subcomponents usually belong to the same mechanical structure, it is actually unknown which one will be most affected by the initial impulse. Second, let us assume that the initial perturbation is such that the energy flow takes the desired direction. Then, according to the Poincaré recurrence theorem, at some point, the energy flow will reverse its direction unless permanently harvested from the absorbing unit or dissipated before the next half wave of the beating cycle takes place. Some alternative approaches suggest oscillators with nonsmooth and discontinuous restoring force characteristics as energy absorbers [18–20]. The advantage is that such type of oscillators can interact with the main structure in a much wider range of frequencies. As shown in [21], a passive energy sink can be created even without damping units based on the property of chaotic soft-wall billiards. In this case, the trend in the energy flow will be statistically one-directional from the main structure towards to billiard leading to the increase of kinetic energy of the inner particle(s). Assuming no damping inside the billiard, the recurrence still can occur, however, it is unlikely to happen within any physically reasonable operational time of the system. Studies contained in [22] present a system of a truss core sandwich beam with a nonlinear energy sink (NES) mounted inside in order to suppress its vibrations. Vibration damping is based on the principle of nonlinear targeted energy transfer. The cases of vibrations caused by impulse and harmonic loads were analyzed. It has been shown that the damping can be enhanced by increasing the mass of NES and locating it in the midpoint of the beam.

Moreover, energy transfer between the components of the dynamical system can be observed in the vibration of composite laminated plates. The nonlinear dynamics of this type of systems was reported in [23]. Experimental, analytical and numerical studies indicate the occurrence of jump phenomena as well as hardening and softening resonances. Due to the nonlinearity of the system, energy can be transferred from the small-amplitude high frequency components of the motion to the large amplitude low frequency components of the motion [24]. Furthermore, a robust control method is proposed in [25] for the vibration suppression of the piezoelectric laminated composite cantilever rectangular plate subjected to the aerodynamic force in the hygrothermal environment. Piezoelectric layers work in this system as both a sensor and an actuator. Numerical tests were carried out for various parameters of temperature, moisture concentration, aspect ratio, and damping. It has been confirmed that the active vibration control significantly suppresses the vibration amplitude compared to the uncontrolled system.

The dynamics of coupled pendulums systems and other multi-degree-oscillators subjected to an alternating magnetic field has already been studied by many scientists. The problem of excitation and synchronization of vibrations in a complex system of two coupled magnetic pendulums was analyzed using the control speed-gradient method [26,27]. The objective function was designed as a weighted sum of two components in order to achieve two goals, such as maintaining the certain value of total energy level and the antiphase mode of vibrations. It has also been demonstrated that the excitability index [28], which measures the resonance properties of systems, can be calculated by an approximated fast gradient method giving a reasonable accuracy for real systems. Nonlinear effects occurring in the system of coupled magnetic pendulums are considered in [29–31]. In particular, reference [29] deals with numerical studies of the bifurcation dynamics with rich physical effects in both regular and chaotic motions. The existence of multi-periodic, quasi-periodic and chaotic solutions was shown numerically and experimentally. The case of quasi-periodicity was confirmed with the help of Lyapunov exponents, while the basins of attraction computed for this case showed the existence of two other periodic solutions depending on the initial values. Chaotic attractors with their neighboring periodic solutions and coexisting of the regular attractors have been reported in [30]. The magnetic field has a significant effect on the potential energy by generating multi-wells responsible for the complex dynamics analyzed in [32]. Resonant phenomena and periodic solutions were investigated analytically in reference [31]. The harmonic balance method was applied to approximate subharmonic and quasi harmonic regimes. More complex motions were explored by a successive maxima map. In particular, it was observed that the transition to chaos was not accompanied by cascades of periodic doubling. Recently, two magnetically coupled oscillators were tested for absorbing energy by one them from the other [33]. The main goal was to employ the saturation phenomena of the 1:3 internal resonance and investigate the effect of reducing the amplitude of the primary resonance of one of the oscillators. One finding is that energy transfer between oscillators can take place through harmonic frequencies generated by magnetic nonlinearity. A significant influence of the nonlinearity of magnetic interactions on the dynamics of a 2-DoF system, made of coupled oscillators with magnetic springs, was also demonstrated in [34]. Different mathematical models of this system were considered. It was shown that a cascade of doubling bifurcation, chaotic attractors, hysteresis behavior, amplitude jumps, and quasi-periodicity may develop in the system dynamics. Similar phenomena were also observed in paper [35].

A large group of scientists utilize coupled magnetic oscillators, including pendulums as energy harvesting devices [36–41], describing their advantages and disadvantages compared to other energy harvesters.

As follows from this brief literature overview, the concept of targeted energy transfer in physical systems has been under investigation for more than two decades. A significant attention was paid to the passive mechanisms of guiding the energy flows typically from the main structure to a smaller substructure including a dissipative device. In contrast, the present work develops a methodology for guiding the energy flow in a prescribed direction between two identical structural elements. A novelty of the approach is in physical principles of actuation provided by controllable electromagnetic fields surrounding the vibrating structure. It is shown at the preliminary stage of the study that the magnetic fields can effectively change the system parameters and thus create a necessary time dependent asymmetry between two physically identical oscillators. The corresponding analysis leads to a specific feedback relationship, which is used for calculation of electromagnet currents. A reason for considering the couple of identical oscillators is that such a study can provide some basis for further extensions of the methodology on one-dimensional continuous structures and their discretized models with the translational spatial symmetry. For that reason, two coupled pendulums represent a convenient model facilitating the experimental setup and illustrating basic features of the approach.

This paper is organized as follows. Section 2 focuses on the experimental setup construction. Section 3 describes mathematical modeling, the goal of study, and some preliminary remarks. Section 4 deals with analytical methodology including basic manipulations with the differential equations of motion, such as transition to new descriptive quantities and averaging procedure. The developed approach is validated based on a comparison of the response of reduced system with the results of direct numerical integration of the original model. Section 5 is devoted to the model validation based on comparison with experimental measurements. In particular, the effectiveness of an open-loop control strategy is discussed. Section 6 deals with a qualitative discussion of the reduced system in order to develop a closed loop control strategy, which is further implemented in Section 7. Finally, Section 8 provides concluding remarks.

## 2. Experimental setup

In this section, the experimental setup, on which we performed the empirical investigations contained in the work, is presented. The experimental setup is shown in Fig. 1a and consists of two pendulums 1-2, which are made of aluminum and textolite materials. Neodymium magnets 3 are mounted at the end of both pendulum arms (they are not directly visible in the figure). During experimental studies, we used two pairs of magnets with different diameters. Detailed technical data of components used in the experimental rig are presented in Table 1. The pendulums are mounted on brass shafts 4 supported by ball bearings that holders are attached to an aluminum frame of the setup. The shafts are coupled with a steel torsion spring 7. Since the pendulums are weakly coupled, the low value of spring stiffness (parameter in Table 2) has been achieved by a small number of coils and wire diameter, and a large outer diameter. Electric coils 5-6 are placed under the pendulums. The coils are fed with the current signals from the R&S®NGL202 power supply, which is controlled by a PC. The magnet-coil components are similar in concept to a classic electromagnetic transducer, i.e., depending on the direction of the current flowing in the coil, the magnet may be repelled or attracted by the coil. The magnetic interaction between these components depends mainly on the value of the current, but it is also dependent on the external dimensions of the magnets and the coil including the initial clearance between their surfaces when the pendulum is at rest. An increase in the dimensions strengthens their interaction, while an increase in the initial distance weakens this interaction. The constant values of these dimensions the size of the gap are hidden in the system parameters *a* and *b* in Table 2. Since the used power supply cannot change the direction of the coil current flow, we employ a specially designed and built electronic relay system operating on the principle of an H-bridge. In our studies, the positive value of the current signal generated the magnetic field repulsing the magnet, while the negative value of the current was responsible for the attraction of the magnet.

Table 1. Properties of magnets, electric coil and torsion spring used in the experimental setup.

| Torsion spring | | |
|---|---|---|
| Parameter | Value | Unit |
| Outer diameter | 57 | *mm* |
| Wire diameter | 0.6 | *mm* |
| Total coils | 4 | - |
| **Electric coil** | | |
| Parameter | Value | Unit |
| Outer diameter | 40 | *mm* |
| Resistance | 10.9 | *Ω* |
| Inductance | 22 | *mH* |
| Initial distance from the magnet | 23 | *mm* |

| Large magnets | | | Small magnets | | |
|---|---|---|---|---|---|
| Parameter | Value | Unit | Parameter | Value | Unit |
| Diameter | 22 | *mm* | Diameter | 14 | *mm* |
| Height | 10 | *mm* | Height | 10 | *mm* |
| Mass | 28.36 | *g* | Mass | 11.42 | *g* |

Measurements of the pendulum angular positions are carried out with the use of optical sensors HEDS-9040#J00 with a resolution of 0.36 degrees. The signals from the sensors as well as the coil currents are collected by the data acquisition card NI USB-6341 and processed by the PC. The schematic flow of signals between the elements of the setup is shown in Fig. 1b. In conducted experiments there were two cases of controlling the pendulums using magnetic field generated by the coils:

i). Only pendulum 1 was subjected to a magnetic field generated by coil 5. In this case coil 5 works in a short circuit with a power supply, whereas, the electrical circuit of coil 6 was open at that time. The possibility of the open circuit of coil 6 is depicted by the dashed open switch in Fig. 1b;

ii). Both pendulums 1 and 2 were subjected to a magnetic field generated by both coils 5 and 6. In this case both coils work in a short circuit with a power supply.

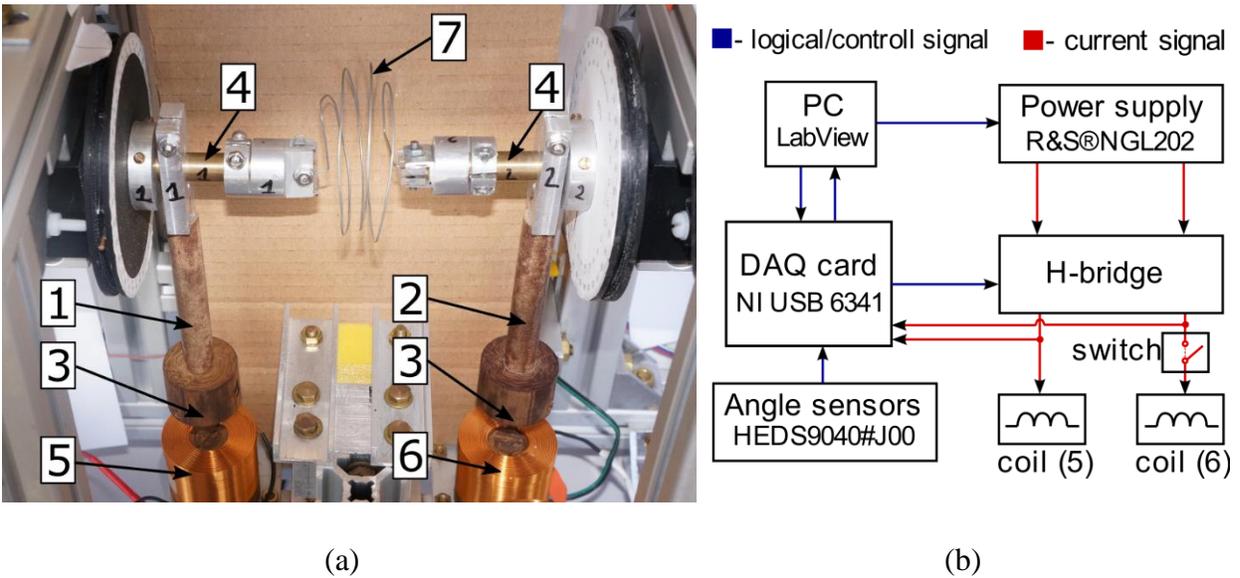

(a) (b)

Figure 1: (a) Experimental setup of weakly coupled pendulums subjected to a magnetic field, where the labeled components are as follows: 1, 2 – pendulums, 3 – neodymium magnets, 4 – brass shafts, 5, 6 – electric coils, 7 – steel torsion spring, and (b) the diagram of signal flows in the experimental setup.

Since the present research concerns guiding the energy flow between pendulums by means of magnetic fields, the coil-magnet components are not a primary source of excitation for the pendulums. These structural components are installed to generate 'steering forces' controlling the energy distribution without affecting its level by varying the system properties. The external energy influx is provided by assigning nonzero initial conditions for the dynamic states, such that the pendulums are deflected from the equilibrium positions and then released. It will be shown in Section 3.2 below that the role of magnetic actuators is to effectively vary parameters of the system rather than changing its total energy level.

## 3. Modeling and problem formulation

### 3.1 *The goal of study and model description*

A general model under study is schematically shown in Fig. 2a, which is a typical two mass-spring system oscillating in a controllable magnetic field due to some initial disturbance. Assuming that both the elastic coupling and dissipation are relatively weak and the system is close to 1:1 resonance condition, some portion of the mechanical energy will be cyclically moving back and forth from one oscillator to another in beat wise manner until the total energy completely dissipates. The goal of the work is to understand how such an energy exchange flow can be controlled with an artificially created magnetic field around the mechanical structure, supposedly the massive elements can interact with the magnetic field. Without loss of generality, with reference to Fig. 1a and Fig. 2b, we consider a system of two compound pendulums weakly coupled with a linear torsional spring of stiffness $k_e$. Each pendulum has its mass $m$ and a moment of inertia $J$ about the axis of rotation. The distance from the axis of rotation to the pendulum mass center is $s$, and $g$ is the gravitational acceleration.

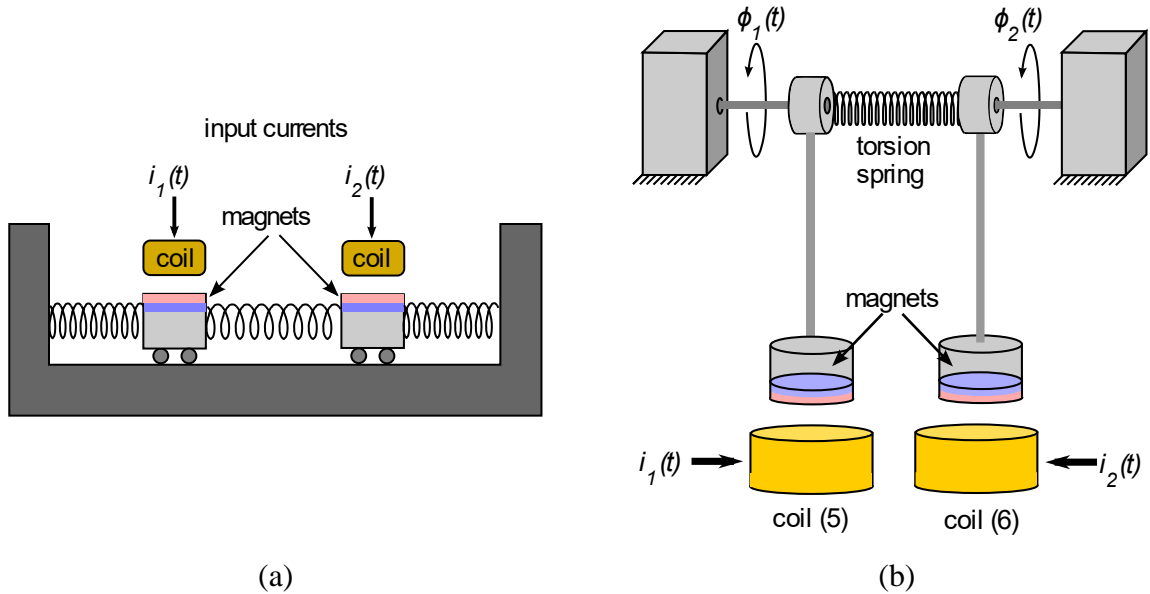

(a)          (b)

Figure 2: Weakly coupled compound pendulums subjected to electromagnetic loading: (a) schematic diagram of a general two mass-spring system in a controllable magnetic field, and (b) schematic model of the experimental rig – two pendulums coupled with a torsion spring and subjected to both gravitational and magnetic fields.

Based on the experimental data, it was found that the dissipative effects in bearings admit approximation with Coulomb dry friction law characterized by the coefficients $c_1$ and $c_2$ for each pendulum, respectively. Furthermore, the coupling spring is assumed to have a structural (material) damping with the equivalent linear damping coefficient $c_e$. Also, due to the goal of our study, it is sufficient to consider relatively small angular displacements $\phi_i$ of both pendulums, such that $|\phi_i| << \pi/2$ ($i = 1, 2$), with a cubic approximation of the gravitational restoring torque characteristics. Based on the introduced notations and assumptions, the governing differential equations of motion are represented in the form

$$J\ddot{\phi}_1 = -c_1 \operatorname{sgn}\dot{\phi}_1 - c_e(\dot{\phi}_1 - \dot{\phi}_2) - mgs\left(\phi_1 - \frac{1}{6}\phi_1^3\right) - k_e(\phi_1 - \phi_2) + i_1(t)\frac{dU_{mag}(\phi_1)}{d\phi_1},$$
$$J\ddot{\phi}_2 = -c_2 \operatorname{sgn}\dot{\phi}_2 - c_e(\dot{\phi}_2 - \dot{\phi}_1) - mgs\left(\phi_2 - \frac{1}{6}\phi_2^3\right) - k_e(\phi_2 - \phi_1) + i_2(t)\frac{dU_{mag}(\phi_2)}{d\phi_2}, \quad (1)$$

where $i_1(t)$ and $i_2(t)$ are input currents into the electric coils as shown in Fig. 2b, and $U_{mag}(\varphi)$ is a magnetic potential per unit current, which is approximated in a phenomenological way as [32]

$$U_{mag}(\varphi) = a\left[1 - \exp\left(-\frac{\varphi^2}{b}\right)\right]. \quad (2)$$

Here $\varphi$ denotes the angular displacement, while $a$ and $b$ are constant parameters characterizing physical properties and geometry of each coil-magnet pair. According to equations (1), the potential given by relationship (2) generates a magnetic torque per unit current as

$$M_{mag}(\varphi) = \frac{dU_{mag}(\varphi)}{d\varphi} = \frac{2a}{b}\exp\left(-\frac{\varphi^2}{b}\right)\varphi. \quad (3)$$

Both dependencies (2) and (3) are illustrated by Fig. 3. In our experiments, two types of magnet pairs that differ in size are used. The corresponding sets of parameters were obtained by fitting experimental data and listed in Table 2. However, both types of magnets appeared to have qualitatively similar potentials and moments as shown in Fig. 3.

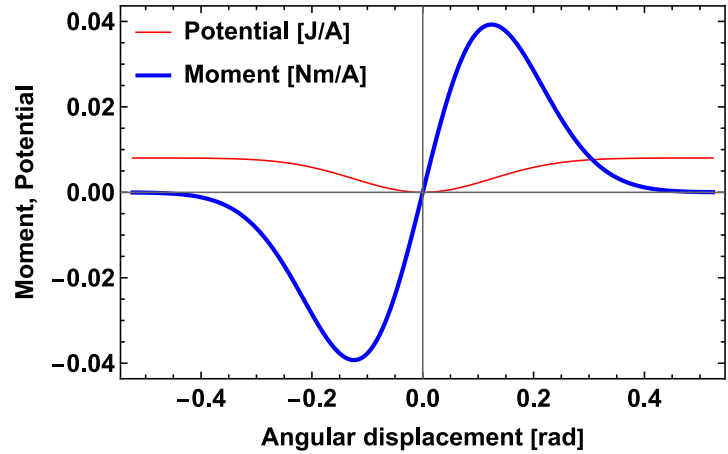

Figure 3: Profiles of the magnetic potential $U_{mag}(\varphi)$ and the corresponding torque per unit current, $M_{mag}(\varphi)$, obtained for the numerical values of parameters $a$ and $b$ from Table 2 (Large magnets).

Regarding the adapted empirical approximation for the potential $U_{mag}$ note that, a typical assumption of the point-wise magnet-coil interaction [27, 28] is rather unapplicable in our case due to a large size of interacting bodies compared to distances between them; see Fig. 1a. An alternative consideration of three-dimensional specifics of a magnet-coil pair would require technically complicated procedures [44] shadowing the main goal of the present study. Therefore, using experimental curves depicting typical features of the magnetic interaction, as shown in Fig.3, represents a reasonable option.

Table 2. System parameters identified from experimental data for a pair of large and small magnets.

|  | Large magnets |  |  | Small magnets |  |
|---|---|---|---|---|---|
| Parameter | Value | Unit | Parameter | Value | Unit |
| $a$ | $8.036 \cdot 10^{-3}$ | $Nm \cdot rad \cdot A^{-1}$ | $a$ | $3.635 \cdot 10^{-3}$ | $Nm \cdot rad \cdot A^{-1}$ |
| $b$ | $30.810 \cdot 10^{-3}$ | $rad^2$ | $b$ | $43.366 \cdot 10^{-3}$ | $rad^2$ |
| $c_1$ | $2.500 \cdot 10^{-4}$ | $Nm$ | $c_1$ | $3.114 \cdot 10^{-4}$ | $Nm$ |
| $c_2$ | $1.600 \cdot 10^{-4}$ | $Nm$ | $c_2$ | $2.705 \cdot 10^{-4}$ | $Nm$ |
| $c_e$ | $9.615 \cdot 10^{-6}$ | $Nms \cdot rad^{-1}$ | $c_e$ | $9.615 \cdot 10^{-6}$ | $Nms \cdot rad^{-1}$ |
| $k_e$ | $3.999 \cdot 10^{-3}$ | $Nm \cdot rad^{-1}$ | $k_e$ | $3.999 \cdot 10^{-3}$ | $Nm \cdot rad^{-1}$ |
| $J$ | $6.787 \cdot 10^{-4}$ | $kg \cdot m^2$ | $J$ | $5.675 \cdot 10^{-4}$ | $kg \cdot m^2$ |
| $mgs$ | $5.840 \cdot 10^{-2}$ | $Nm$ | $mgs$ | $5.018 \cdot 10^{-2}$ | $Nm$ |

Recall that Coulomb dry friction law, which is assumed by equations (1), is in agreement with the experimental results showing almost linear amplitude decays; for an overview of experimental friction characteristics see reference [32]. However, the presence of step-wise discontinuous functions of velocities, or even their smooth substitutes, may contradict the so-called Lipschitz condition of existence and uniqueness of solutions. As a result, the phase trajectory can be trapped by the discontinuity surface at least for some time. In physical terms, one of the pendulums can briefly stop until another pendulum develops torque in the coupling spring for pooling the first pendulum from its sticking state. Although such effects may be barely seen in the time history graphs, a detailed analysis shows the possibility of qualitative changes such that inphase oscillations switch to antiphase or vice versa. Fortunately, this is likely to occur at low excitation levels as follows from the results of simulations described below.

### 3.2 Preliminaries

In order to provide a preliminary insight into the physical mechanisms of control inputs, let us assume no damping and keep only quadratic terms of the polynomial expansion for the system potential energy combining both the effect of gravity and magnetic field as

$$V(\phi_1, \phi_2) = \frac{1}{2} \Omega^2 \left[ \phi_1^2 + \beta(\phi_1 - \phi_2)^2 + \phi_2^2 \right] - \frac{a}{bJ}(i_1 \phi_1^2 + i_2 \phi_2^2), \quad (4)$$

where $\Omega = \sqrt{mgs/J}$ is a natural frequency of the linearized pendulums, $\beta = k_e / mgs$ is a relative strength of coupling; other parameters and their values are listed in Table 2 and Table 3.

This potential energy function corresponds to a linearized conservative system. Note that the quadratic approximation for the magnetic part in Eq. (4) is used only for the discussion in this section. Further the magnetic torques will be considered in the original form (see Fig. 3) described by relationship (3). Since our control strategies require a relatively slow variation of currents compared to the temporal rate of oscillations, both values, $i_1$ and $i_2$, can be viewed as parameters changing the shape of the potential surface in an adiabatic way. While the presence of magnetic field has no effect on the location of the stationary point, $(\phi_1,\phi_2)=(0,0)$, its type can be affected by the currents. The diagram in Fig. 4 illustrates how the level lines of the potential energy, $V(\phi_1,\phi_2)=const.$, evolute with currents in the neighborhood of the stationary point (0,0). The elliptic shapes from the left bottom part of the diagram confirm that negative currents preserve the minimum of potential energy and therefore maintain Lyapunov's stability of the equilibrium at zero. In contrast, increasing either one or both currents eventually transforms the level lines of potential surface by indicating its switch from concave to convex or to a saddle, respectively. This means that the repelling magnetic loads become dominating over the coupling and gravitational restoring torques.

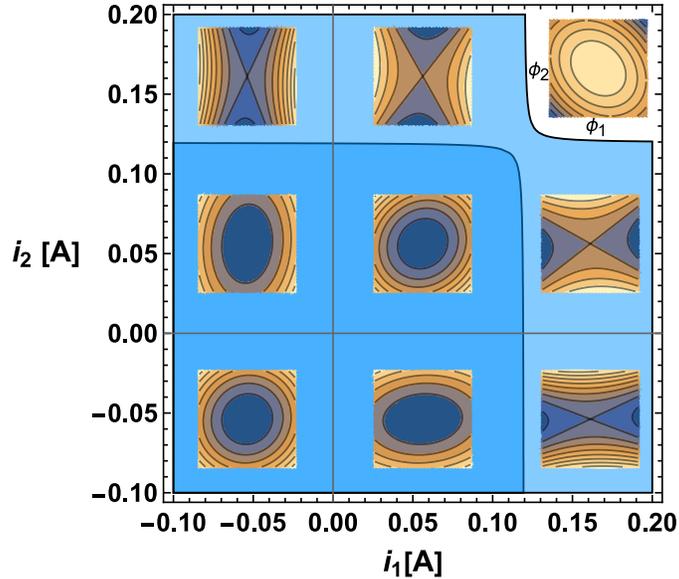

Figure 4: Evolution of the level lines of the effective potential energy (4) along different intervals of the magnet currents: the square dark blue area of elliptic shapes refers to a minimum of the potential surface, the light blue area represents a saddle, and the white area at the upper right corner corresponds to a maximum; the boundaries between different areas are obtained by sequentially setting to zero the natural frequencies of linearized model given by Eq. (5) for the parameters listed in Table 2 for the case of large magnets. (Color online.)

Quantitative estimates for the corresponding critical magnitudes of currents can be obtained in a standard algebraic way by analyzing the quadratic form (4) through the Jacobian of function $V(\phi_1,\phi_2)$ or by considering the eigen frequencies of the linearized vibrating system assuming the currents to be fixed:

$$\omega_{1,2}^2 = (1+\beta)\Omega^2 - \frac{a}{bJ}(i_1+i_2) \mp \sqrt{\beta^2\Omega^4 + \frac{a^2}{b^2J^2}(i_1-i_2)^2} \ . \tag{5}$$

Obviously, both frequencies must be real in the left bottom area of the diagram (see Fig. 4), where function (4) has minimum. Let us note that the situation may appear to be quite different in the case of time-varying currents in temporal rates comparable with the natural temporal rates of pendulums, when the effects of parametric resonance are possible.

To conclude this section, Eqs. (1), (4), and (5) reveal that both currents have a parametric effect on the system by effectively varying the stiffness terms of the linearized system. Therefore, considering the currents as control inputs is equivalent to the intent to control the system dynamics by manipulating the system properties rather than external loading. In particular, Eq. (5) shows that varying the currents in a relatively slow adiabatic way can shift the pendulums either from or towards the 1:1 resonance condition in order to achieve a desired trend in the energy exchange process.

## 4. Analytical procedure

### 4.1 *Model adaptation*

Our analytical approach to the control strategies and interpretations of the experimental data is based on the idea of averaging the system with respect to a single fast phase as soon as both linearized pendulums have the same natural frequency. This is a necessary assumption for considering a 1:1 resonance energy flow. Any frequency detuning should be viewed as a small structural perturbation of the generating system. Developing control algorithms for the energy flows, requires appropriate transition from the original state variables to the energy related descriptive variables, and then averaging over the fast oscillations. These steps are described below in the present section. First, we represent system (1) in the form of four first-order equations for the set of state variables including angular displacements $\phi_j$ and angular velocities $v_j$ of the pendulums as

$$\frac{d\phi_j}{dt} = v_j, \quad \frac{dv_j}{dt} = -\Omega^2 \phi_j - f_j; \quad j = 1, 2, \tag{6}$$

where

$$f_1 = 2\Omega\left[\zeta_1 \operatorname{sgn} \dot{\phi}_1 + \alpha(\dot{\phi}_1 - \dot{\phi}_2)\right] + \Omega^2\left[\beta(\phi_1 - \phi_2) - \frac{1}{6}\phi_1^3\right] - \frac{i_1(t)}{J}\frac{dU_{mag}(\phi_1)}{d\phi_1},$$

$$f_2 = 2\Omega\left[\zeta_2 \operatorname{sgn} \dot{\phi}_2 + \alpha(\dot{\phi}_2 - \dot{\phi}_1)\right] + \Omega^2\left[\beta(\phi_2 - \phi_1) - \frac{1}{6}\phi_2^3\right] - \frac{i_2(t)}{J}\frac{dU_{mag}(\phi_2)}{d\phi_2}, \tag{7}$$

and the following unitless parameters are used:

$$\Omega = \sqrt{\frac{mgs}{J}}, \quad \beta = \frac{k_e}{mgs}, \quad \zeta_1 = \frac{c_1}{2J\Omega}, \quad \zeta_2 = \frac{c_2}{2J\Omega}, \quad \alpha = \frac{c_e}{2J\Omega}. \tag{8}$$

According to the data of Table 2, numerical values of parameters (8) for two cases under consideration are given in Table 3.

Table 3. Unitless parameters of a system equipped with a pair of large and small magnets.

| Large magnets | | Small magnets | |
|---|---|---|---|
| Parameter | Value | Parameter | Value |
| $\Omega$ | 9.276 | $\Omega$ | 9.403 |
| $\beta$ | $6.849 \cdot 10^{-2}$ | $\beta$ | $7.969 \cdot 10^{-2}$ |
| $\zeta_1$ | $1.985 \cdot 10^{-2}$ | $\zeta_1$ | $2.918 \cdot 10^{-2}$ |
| $\zeta_2$ | $1.270 \cdot 10^{-2}$ | $\zeta_2$ | $2.534 \cdot 10^{-2}$ |
| $\alpha$ | $7.636 \cdot 10^{-4}$ | $\alpha$ | $9.009 \cdot 10^{-4}$ |

Note that the present damping ratios $\zeta_i$ $(i=1, 2)$ do not characterize the damping effect in similar to the linear theory way, since the dry friction model is assumed here. However, replacing $\mathrm{sgn}\,\dot\phi_i$ with $\dot\phi_i$ would provide these damping ratios with the conventional meaning whenever the case of viscous damping becomes of interest.

**4.2 *Descriptive functions and the coordinate transformation***

As mentioned at the beginning of Section 4.1, our analytical approach is based on a transition from the system (6)-(7) to a system which is well suited for describing the energy exchange process between two pendulums. The related methodology was introduced earlier for the analyses of energy exchange between liquid sloshing modes [45] and beating effects in friction induced vibrations of two weakly coupled oscillators [46]. Basic steps of analytical manipulations are reproduced below. In particular, we show that the excitation levels of pendulums and the phase shift between the pendulums will be characterized through the elements of the symmetric energy matrix

$$E_{kj} = \frac{1}{2}\left(\frac{d\phi_k}{dt}\frac{d\phi_j}{dt} + \Omega^2 \phi_k \phi_j\right), \quad k, j = 1, 2, \tag{9}$$

where $\phi_{k,j}$ ($k, j = 1,2$) are angular displacements of the pendulums, and $\Omega$ is the natural frequency of a linearized pendulum as introduced in Eq. (8).

Neglecting dissipation, nonlinearity, coupling, and also setting currents to zero provides quantities (9) with a clear physical meaning: $E_{11}$ and $E_{22}$ are total energies of both decoupled pendulums, whereas the elements $E_{12} = E_{21}$ appear to carry information about the phase shift. In order to reveal the connection to the phase shift, let us assume that both pendulums are linearized and decoupled and thus described by solutions $\phi_1 = A_1 \cos\Omega t$ and $\phi_2 = A_2 \cos(\Omega t + \Delta)$ with the phase shift $\Delta$. Substituting these solutions in (9) gives

$$E_{12} = \frac{1}{2}\left(\dot\phi_1\dot\phi_2 + \Omega^2\phi_1\phi_2\right) = \frac{1}{2}\Omega^2 A_1 A_2[\sin\Omega t \sin(\Omega t + \Delta) + \cos\Omega t \cos(\Omega t + \Delta)]$$
$$= \frac{1}{2}\Omega^2 A_1 A_2 \cos\Delta \tag{10}$$

and

$$E_{11}E_{22} = \frac{1}{4}\Omega^4 A_1^2 A_2^2. \qquad (11)$$

The phase shit $\Delta$ is therefore determined by the ratio

$$\frac{E_{12}}{\sqrt{E_{11}E_{22}}} = \cos\Delta. \qquad (12)$$

The reason for using the quantities $E_{kj}$ as the system descriptive variables is that such variables describe the system in energy units eliminating a less informative fast varying temporal scale $\Omega t$. It will be seen from Eq. (13) below that algebraic combinations of different elements $E_{kj}$ provide a sufficient information about the dynamic states of two identical oscillators in an explicit way. Unfortunately, quantities $E_{11}$ and $E_{22}$ are losing their straightforward physical meaning for the entire system due to the dissipation and other factors. Nevertheless, they still can serve for a characterization of excitation levels of the individual pendulums. From such a standpoint, we continue using the term "energy" and introduce the following combinations of $E_{kj}$:

$$\begin{aligned}
E &= E_{11} + E_{22}, \\
P &= \frac{E_{11} - E_{22}}{E}, \quad -1 \le P \le 1, \\
Q &= \frac{E_{12}}{\sqrt{E_{11}E_{22}}} = \cos\Delta, \quad -1 \le Q \le 1,
\end{aligned} \qquad (13)$$

where $E$ is the total energy of the oscillators, $P$ is the energy partitioning between the oscillators, and $Q$ stands for the 'coherency index' characterizing the phase shift $\Delta$ between the oscillators. In particular, the number $P=0$ means that the energy is distributed equally between the pendulums. The number $P=1$ points out that only the first pendulum is oscillating, whereas the second one remains at rest. When $P=-1$ then only the second pendulum is swinging. The coherency index $Q$ determines the type of synchronous mode as follows: $Q=-1$ ($\Delta=\pi$) – antiphase mode, $Q=0$ ($\Delta=\pi/2$) – the elliptic mode, and $Q=1$ ($\Delta=0$) – the inphase mode. A detailed geometrical interpretation of this classification is given in [46]. Further, we also use the explicit phase shift parameter $\Delta$ in transformation from (13) to the original state variables $\{E, P, \Delta, \delta\} \to \{\phi_1, v_1, \phi_2, v_2\}$. In order to obtain explicit expressions for such a transformation, let us substitute $\phi_1 = A_1 \cos\Omega t$ and $\phi_2 = A_2 \cos(\Omega t + \Delta)$ in Eq. (9) and then Eq. (13):

$$\begin{aligned}
E &= E_{11} + E_{22} = \frac{1}{2}\Omega^2\left(A_1^2 + A_2^2\right), \\
P &= \frac{E_{11} - E_{22}}{E} = \frac{1}{2E}\Omega^2\left(A_1^2 - A_2^2\right).
\end{aligned} \qquad (14)$$

Solving these equations for the amplitudes $A_1$ and $A_2$ finally defines the transformation $\{E, P, \Delta, \delta\} \to \{\phi_1, v_1, \phi_2, v_2\}$:

$$\phi_1 = \Omega^{-1}\sqrt{E(1+P)}\cos\delta,$$
$$v_1 = -\sqrt{E(1+P)}\sin\delta,$$
$$\phi_2 = \Omega^{-1}\sqrt{E(1-P)}\cos(\delta+\Delta), \quad (15)$$
$$v_2 = -\sqrt{E(1-P)}\sin(\delta+\Delta),$$

where the phase $\Omega t$ is replaced with a general fast phase $\delta = \delta(t)$.

Recall that transformation (15) was designed by using solutions of two independent harmonic oscillators, when the quantities $E$, $P$, and $\Delta$ are fixed, and $\delta = \Omega t$. In general case, the oscillators are coupled and subjected to different kind of perturbations. Nonetheless, under conditions that the coupling is relatively weak and perturbations are small, transformation (15) still can be applied following the idea of parameter variations and assuming that $E$, $P$, and $\Delta$ are not constant any more but slowly varying quantities.

### 4.3 *Averaging procedure*

The averaging of system (6) can be now conducted analogously to the van der Pol averaging. The corresponding procedure includes the substitution of Eqs. (15) in Eqs. (6), then solving the equations for the derivatives $d\{E,P,\Delta,\delta\}/dt$, and finally applying the operator of averaging with respect to the fast phase $\delta$ to the right-hand side of the equations as

$$<\cdots> = \frac{1}{2\pi}\int_0^{2\pi}\cdots d\delta. \quad (16)$$

In the present case, the averaging procedure is justified by both the physical nature of phenomenon under study and assumptions imposed on the model parameters. The averaging is applied to the high frequency carrying oscillations in order to explicitly describe the relatively slow amplitude modulation and phase shift characterizing a gradual energy exchange between the two pendulums under the condition close to 1:1 resonance. For that reason, the external inputs, dissipative effects, coupling, and nonlinearity described with functions $f_i$ ($i = 1,2$) are assumed to be relatively weak compared to the restoring torques in system (6)-(7). The role of generating system is therefore played by the two identical uncoupled harmonic oscillators. In those cases, when a high-order averaging is required, parameters of functions (7) can be re-scaled to prescribe a common small factor to both functions $f_1$ and $f_2$ followed by a procedure of asymptotic integration. However, for most practical reasons considered in physical literature, just the leading-order averaging is applied. Note that the step-wise discontinuities of dry friction violate the so-called Lipschitz condition and thus require additional justification [47,48]. Generally, the system phase trajectory should pass through the discontinuity surface transversely without sliding along this surface; see also the discussion at the end of Section 3.1. In the present case, it may happen near the equilibrium states at very low angular velocities of pendulums. The validity of averaging is confirmed by comparing the results obtained by means of numerical integrations of both the original, (6)-(7), and averaged, (17), systems; see, for instance, Fig. 5d, f, and h. A comparison with experimental results is given below by Fig. 9c and d. Since the comparison is given in terms of quantities $P$ and $Q = \cos\Delta$, Eqs. (9) and (13) are used to process angles and angular velocities obtained from the original system, (6)-(7). The routine and technically complicated analytical manipulations, related to the system transformation and averaging, were conducted in *Mathematica®* package. The final form of the averaged system is given by the following first order differential equations

$$\frac{dE}{dt} = -2\alpha\Omega E\left(1-\sqrt{1-P^2}\cos\Delta\right) - \frac{4}{\pi}\Omega^2\left(\zeta_1\sqrt{\lambda_1}+\zeta_2\sqrt{\lambda_2}\right),$$

$$\frac{dP}{dt} = \Omega\sqrt{1-P^2}(\beta\sin\Delta - 2\alpha P\cos\Delta) - \frac{4}{\pi}\frac{\Omega^2}{E^2}\left(\zeta_1\lambda_2\sqrt{\lambda_1}-\zeta_2\lambda_1\sqrt{\lambda_2}\right),$$

$$\frac{d\Delta}{dt} = \frac{1}{8\Omega}EP - \frac{\Omega}{\sqrt{1-P^2}}(\beta P\cos\Delta + 2\alpha\sin\Delta)$$

$$+ \frac{a}{bJ\Omega}\left\{i_1(t)e^{-\frac{\lambda_1}{2b}}\left[I_0\left(\frac{\lambda_1}{2b}\right) - I_1\left(\frac{\lambda_1}{2b}\right)\right] - i_2(t)e^{-\frac{\lambda_2}{2b}}\left[I_0\left(\frac{\lambda_2}{2b}\right) - I_1\left(\frac{\lambda_2}{2b}\right)\right]\right\},$$

(17)

and

$$\frac{d\delta}{dt} = \left[1 + \frac{1}{2}\beta - \frac{1}{16}\lambda_1 - \frac{1}{2}\sqrt{\frac{\lambda_2}{\lambda_1}}(\beta\cos\Delta - 2\alpha\sin\Delta)\right]\Omega - \frac{a}{bJ\Omega}i_1(t)e^{-\frac{\lambda_1}{2b}}\left[I_0\left(\frac{\lambda_1}{2b}\right) - I_1\left(\frac{\lambda_1}{2b}\right)\right], \quad (18)$$

where $\lambda_1 = E(1+P)/\Omega^2 = 2E_{11}/\Omega^2$ and $\lambda_2 = E(1-P)/\Omega^2 = 2E_{22}/\Omega^2$ are normalized energies of the individual pendulums, and $I_1$, $I_2$ are Bessel functions produced by the averaging of terms related to the magnetic torques. Recall that cubic polynomial approximations are used only for the gravitational torques *mgs* while the magnetic potential and the corresponding torques are kept in the original form (2)-(3) due to their rapid changes near equilibrium positions of the pendulums (Fig. 3).

Equations (17) and (18) provide a complete description of the system dynamics in the leading asymptotic order. The behaviour of fast phase δ is usually not of significant interest. Also, as seen from (18), the temporal behaviour of the fast phase can be estimated separately by the direct integration with respect to time, after equations (17) are solved. Moreover, the right-hand side of Eq. (18) already gives the estimate for the carrying frequency of the process, $d\delta/dt$, which is usually a sufficient information.

### 4.4 *Numerical validation of the reduced model*

In order to validate the averaging procedure leading to Eqs. (17), we compare the results of direct numerical simulations with original Eqs. (6)-(7) to the output obtained from Eqs. (17) in terms of the variables *P* and *Q*. For that reason, numerical solutions of Eqs. (17) for the state variables are processed by means of relationships (9) and (13) to generate the corresponding numerical versions of quantities *P* and *Q*. The simulations were conducted under the input currents

$$i_1(t) = A + B\sin^2\left(\pi\frac{t}{t_k}\right), \quad i_2(t) = -i_1(t), \quad (19)$$

where $A = 0.001$ A, $B = 0.055$ A, and $t_k = 40$ sec.

Equation (19) represents an empirical formula whose parameters are chosen based on a preliminary experimental evaluation of the system physical properties, such as temporal scales, restoring torque magnitudes, etc. For instance, numerical values of *A* and *B* characterize the strength of currents for which the desired effect of energy exchange becomes reasonably visible. Regarding the temporal profile, it is assumed that a current starts at some very low-level *A* – offset, which can always be present, then reaches its maximum, *A+B* at $t = t_k / 2$, and finally drops back to the level *A* at $t = t_k$. The value $t_k$ is simply a natural life time of the process, which depends upon the intensity of dissipation and the initial level of total energy. Equation (19) with different sets of the parameters will be also used in different series of numerical simulations. Recall that positive currents generate repelling magnetic torques whereas negative currents produce attracting torques as follows from the diagram in Fig. 4.

    The results of direct simulations and comparison were obtained for a system equipped with a pair of large magnets and summerized in Fig. 5. The left side of the figure is given in the original notations of equations (6)-(7), whereas the right side gives some insight into the dynamics through the energy distribution, *P*, and coherency, $Q = \cos \Delta$. Also, the right side of Fig.5 confirms a reasonable agreement between the outputs from the original and averaged equations. As follows from Fig. 5, the presence of magnetic fields essentially afects the system dynamics as compared to the response of a free system with almost equally distributed initial energy. In the case of free vibration, the total energy is gradually dissipating while its very small portion keeps moving from one pendulum to another in a beat-wise manner. In contrast, the gradually increasing currents eventually break the symmetry in such a way that the energy is almost completely transferred from the pendulum under repelling magnetic torque to the pendulum, which is oscillating in the attracting magnetic field. This is most clear from Fig. 5b, which is showing the initial equipartition of the energy, $P \sim 0$, is not maintained as the input currents are increasing. Once the input currents have reached a certain level, the energy starts flowing to the pendulum within the attracting magnetic field created by the negative current. At about $t = 12$ sec, the energy distribution takes the lower boundary value $P = (E_{11} - E_{22})/(E_{11} + E_{22}) = -1$, as the second pendulum absorbed all the energy. It should be noticed that, during the entire process, the coherency index remains negative, $Q = \cos \Delta < 0$, revealing that such a one-directional energy flow associates with predominant antiphase oscillations. The corresponding modal contents of the process is illustrated by the trajectory on configuration plane $\phi_1 \phi_2$ (Fig. 5g) and can be characterized as a switch from the antiphase to the localized nonlinear normal mode through a transient 'elliptic mode', $Q \sim 0$, ($\Delta \sim \pi/2$). These remarks will further serve as a basis for developing our control strategies. Note that practically it is important to control the process until the energy has completely moved to the target oscillator for the first time assuming it can be absorbed or just quickly dissipated there. This allows for further simplifications of the system (17) by a linearization with respect to the variable *P* as described in Section 6 and validated here by Fig. 5d.

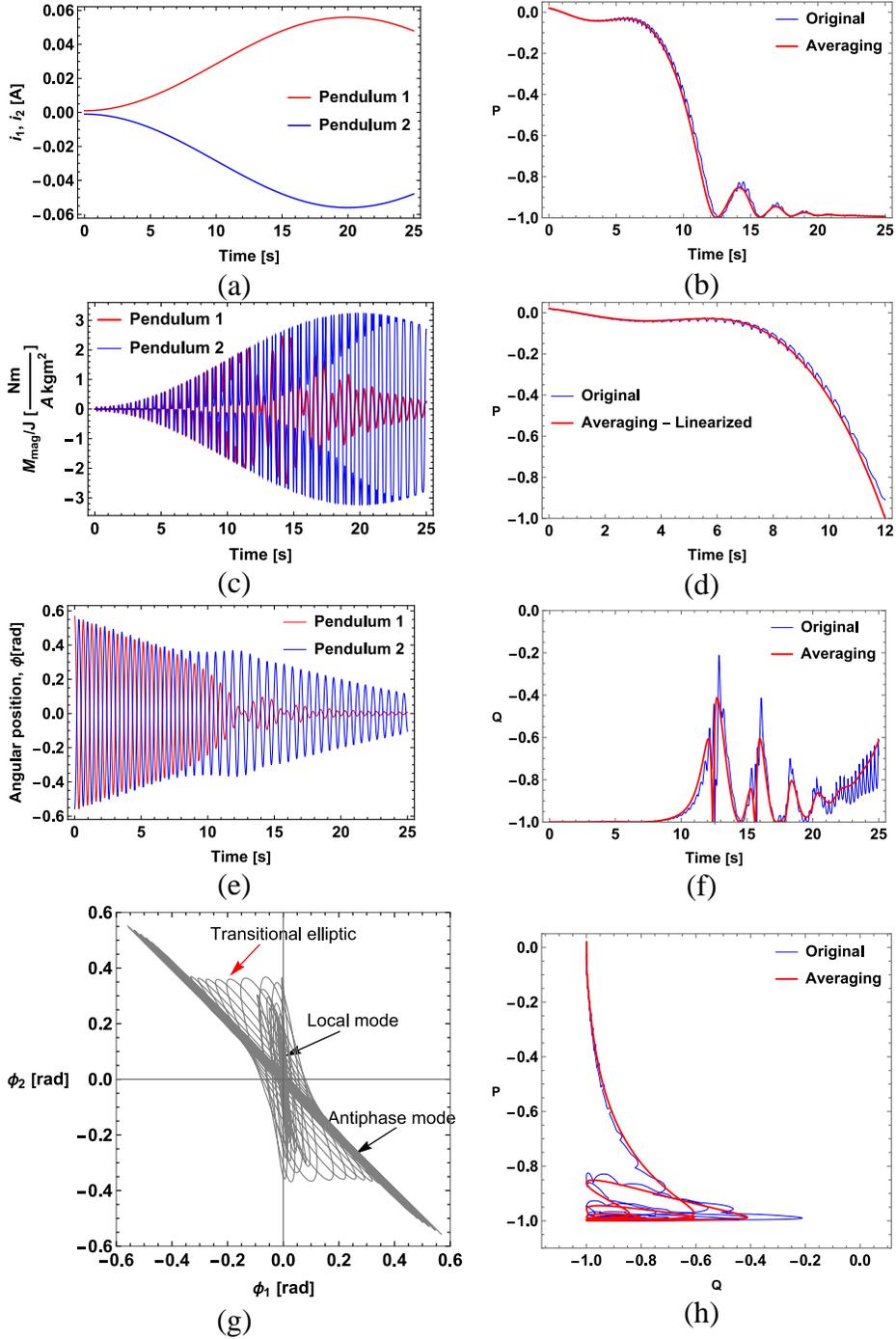

Figure 5: The results of numerical simulations with equations (6)-(7) and (17) under parameters for a pair of large magnets and input currents (19). On the left: (a) profiles of input currents, (c) time histories of magnetic torques, (e) pendulum angular displacements, and (g) transition from the antiphase mode to the localized mode on the plane of configurations. On the right: (b) and (d) time history of the energy distribution, (f) coherency index, and (h) energy distribution versus coherency index. Initial conditions are listed in Appendix A.

### 4.5 *Qualitative analysis of streamlines in the PΔ plane*

The results of Section 4.4, reveal the link between the energy distribution *P* and the modal content of vibrations in terms of the phase shift Δ. Therefore, analyzing the dynamics of system (17) on the plane *P*Δ under the variation of magnetic fields may lead to an adequate control strategy for guiding the energy flow. For that reason, Fig. 6 illustrates the type of stationary points and system trajectories on the plane *P*Δ for different combinations of input currents assuming the total excitation level *E* is 'frozen'. The upper row (a, b, c) of Fig. 6 shows the evolution of streamlines due to the incremental increase of the repelling magnetic field under one pendulum while keeping zero current for another pendulum. The stationary points with $(\Delta, P) = (0,0)$ and $(\Delta, P) = (\pm\pi, 0)$ are associated with the inphase and antiphase vibration modes, respectively; see Eq. (15). Interestingly enough, the dissipation has a destabilizing effect on the stationary points, such that the antiphase modes are always unstable spirals, whereas the inphase mode is changing its type from stable to unstable spiral when the system total energy level drops. Note that such kind of instability means that the structure cannot maintain symmetric mode shapes, which is unrelated to stability properties of the equilibrium configuration.

Further, the bottom row (g, h, i) of Fig. 6 illustrates situations when both pendulums are subjected to magnetic fields produced by currents of the same strength but different combinations of signs. A comparison of these diagrams finally leads to the conclusion that the most effective transition to one of the local modes $P \sim \pm 1$ can be achieved with the currents of opposite signs. It is also clear that input currents of a closed-loop control should depend upon the phase shift Δ. The corresponding feedback dependence will be introduced later in Section 7. It should be emphasized that a gradual dissipation of the total energy does not qualitatively affect the above results provided that the strength of currents is decreased in a proper proportion, although Fig. 6 cannot represent the entire system (17) whose state includes one more variable, *E*, and two time-dependent currents. The phase space of system (17) is therefore four-dimensional: {*E,P,*Δ*,t*}. Still analyzing the diagrams in Fig. 6 explains the link between coil currents and the effect of energy localization on one of the pendulums. A relationship between the topology of streamlines in *P*Δ-plane and the temporal behaviors of values *P* and Δ is clarified below in Fig. 11.

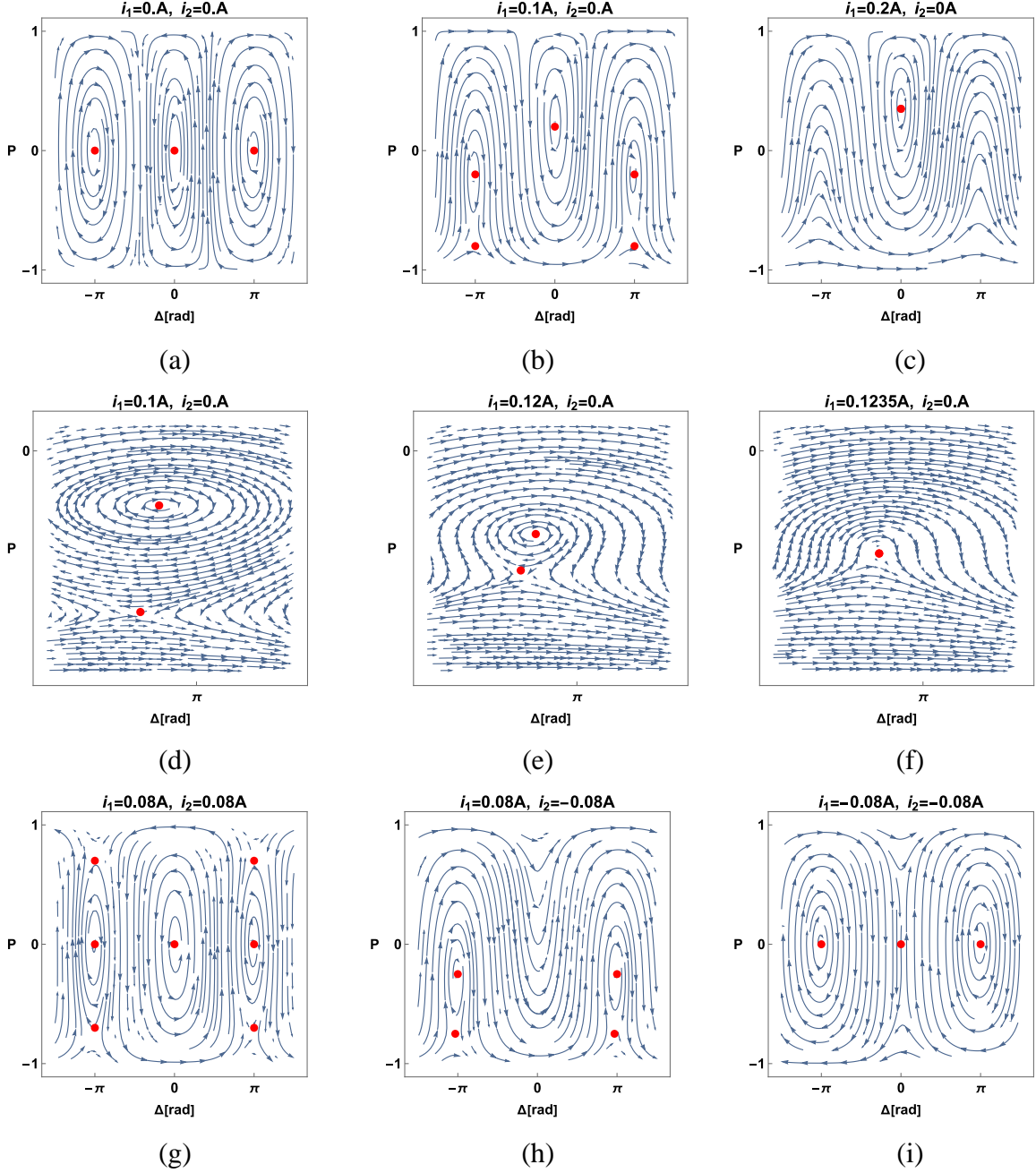

Figure 6: Cross-sections of the phase space of system (17) at fixed excitation level $E = 20$ J and constant coil currents: the first row presents stationary points corresponding to inphase and antiphase modes for (a) free system under zero currents, (b) generation of saddle points corresponding to local modes, and (c) annihilation of the antiphase and local modes producing the 'running phase' $\Delta$; the middle row (d, e, f) illustrates the details of annihilation (c); the bottom row shows (g) the effect of both repelling magnets, (h) combination of repelling and attracting magnets, and (i) both attracting magnets.

## 5. Experimental model validation with open-loop control strategy

As follows from the discussion in Section 4 a real-time information about the phase shift Δ is necessary for an effective closed-loop control of the energy distribution $P$. However, in Section 4.4, a numerical validation of the model was conducted by using the 'open-loop' current inputs (19) showing certain effect on the energy distribution process (Fig. 5b). This happened because relationship (19) incorporated the information about the initial (antiphase) type of oscillations (Fig. 5g). Therefore, as a simpler in implementation, the open-loop control still makes practical sense provided that the initial conditions are a priory known.

### 5.1 *Open-loop control of one pendulum*

Figures 7 and 8 show simulation and experimental time histories of oscillation and energy distributions for an open-loop controlled system. Note that in these cases only pendulum 1 was equipped with a large magnet generating a repelling magnetic field as the coil under the pendulum worked in a short circuit with a power supply, whereas pendulum 2 was not subjected to a magnetic field. Both pendulums were released from different initial angular positions of a predominant antiphase mode with zero angular velocities. Furthermore, Fig. 9 shows the time history of phase shift Δ and the energy distribution $P$ corresponding to the case of Fig. 8.

Overall, the presented time histories show a sufficient qualitative and quantitative match of the experimental records with simulation results.

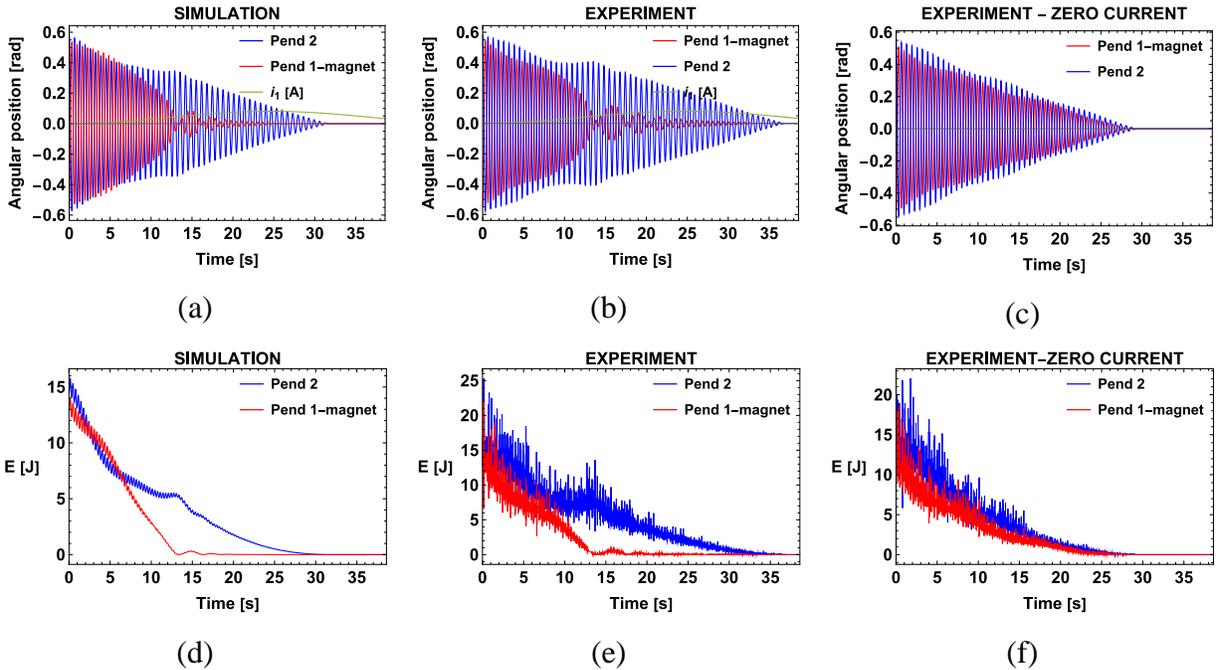

Figure 7: Comparison of results of numerical simulations with experimental records. Time histories of the angular displacements (a)-(c), and the corresponding energy distributions versus time (d)-(f). In cases of (a)-(b) and (d)-(e), the input currents are $i_1(t) = 0.001 + 0.08\sin^2(\pi t / 38.6)$ and $i_2 = 0$, while (c) and (f) illustrate the case of free oscillations under zero input currents $i_1 = i_2 = 0$ for comparison reason. Initial conditions are listed in Appendix A.

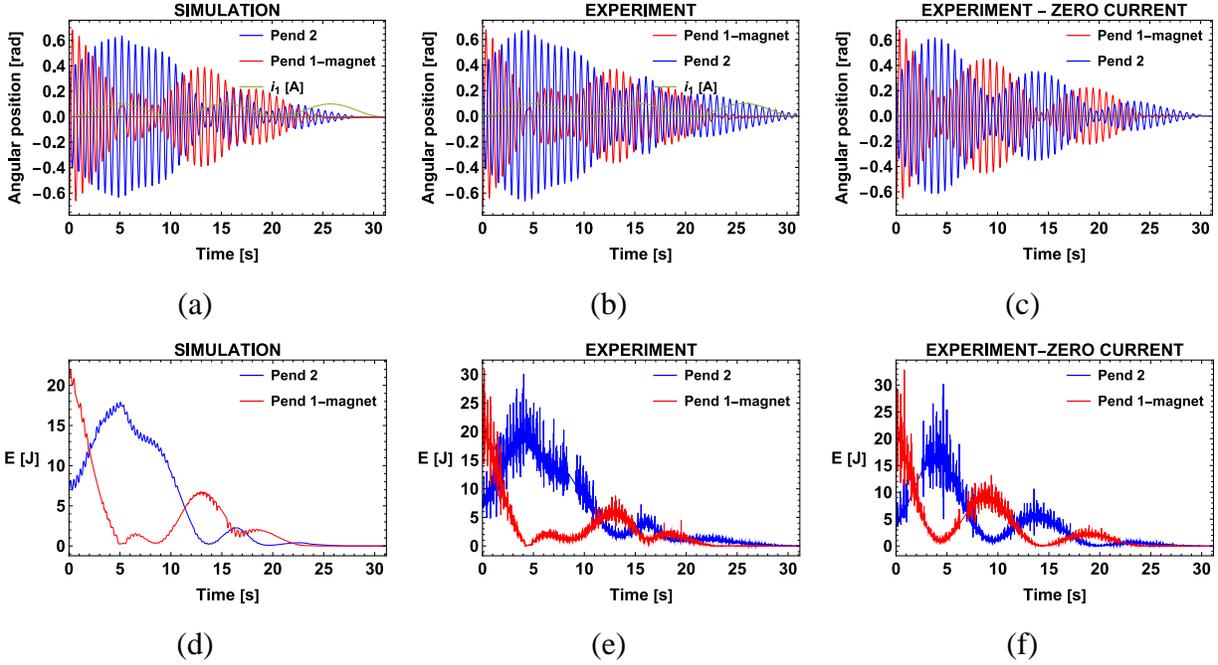

Figure 8: Comparison of results of numerical simulations with experimental records. Time histories of the angular displacements (a)-(c), and the corresponding energy distributions versus time (d)-(f). In cases of (a)-(b) and (d)-(e), the input currents are $i_1(t) = 0.001 + 0.1\sin^2(\pi t / 10.31)$ and $i_2 = 0$, while (c) and (f) illustrate the case of free oscillations under zero input currents $i_1 = i_2 = 0$. Initial conditions are listed in Appendix A.

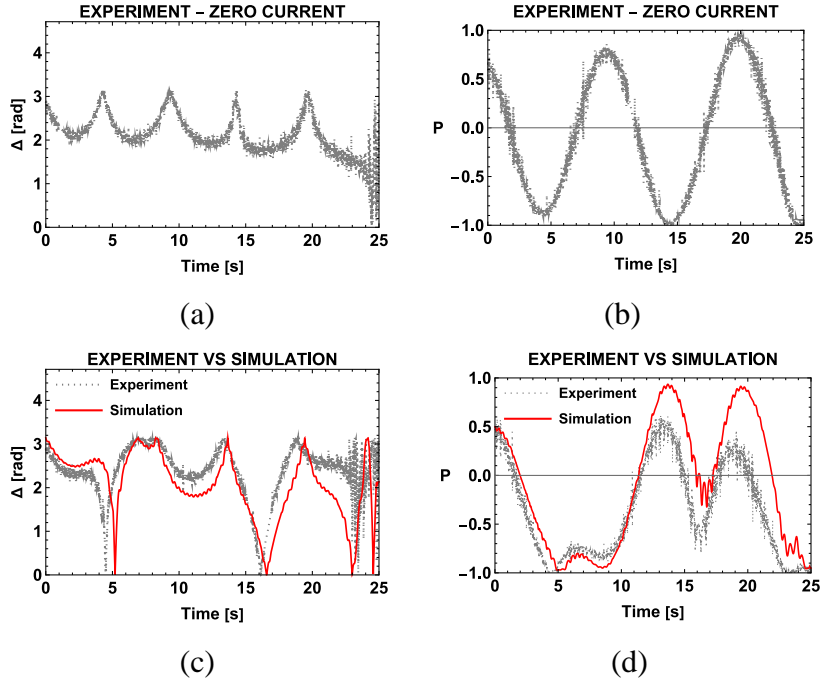

Figure 9: The effect of current inputs in terms of the energy distribution $P$ and phase shift $\Delta$ obtained for results shown in Fig. 8. Time histories (a)-(b) show behaviors of the phase shift and energy distribution under zero current input, while (c)-(d) display the effect of a magnetic field caused by nonzero current input in comparison with numerical simulations based on the original system (6)-(7) and formulas (9)-(13).

## 5.2 *Open-loop control of both pendulums*

The results presented in this subsection were carried out for both pendulums equipped with the small magnets, which means that both pendulums were subjected to a magnetic field as the coils placed under the pendulums worked in short circuits with a power supply. The direction of currents was such that pendulum 1 was subjected to a repelling magnetic field whereas pendulum 2 was under the attracting field. The system was released from its state with different initial angular positions with zero angular velocities. Figure 10 shows simulation and experimental time histories of oscillation and energy distributions of an open-loop controlled system while Fig. 11 illustrates the corresponding behaviors of the energy distribution $P$ and the phase shift $\Delta$. In particular, Fig.11c relates qualitatively different intervals of the phase and the evolution of the system streamlines in the plane $P\Delta$. Since the phase $\Delta$ is bounded by the interval $[0, 2\pi]$, the running phase is represented by the oscillations with the amplitude $2\pi$. Comparing Figs. 11b and 11c reveals that the transition from energy equipartition $(P \sim 0)$ to localization on pendulum 2 $(P \sim -1)$ is accompanied by the transition from antiphase to running phase oscillations.

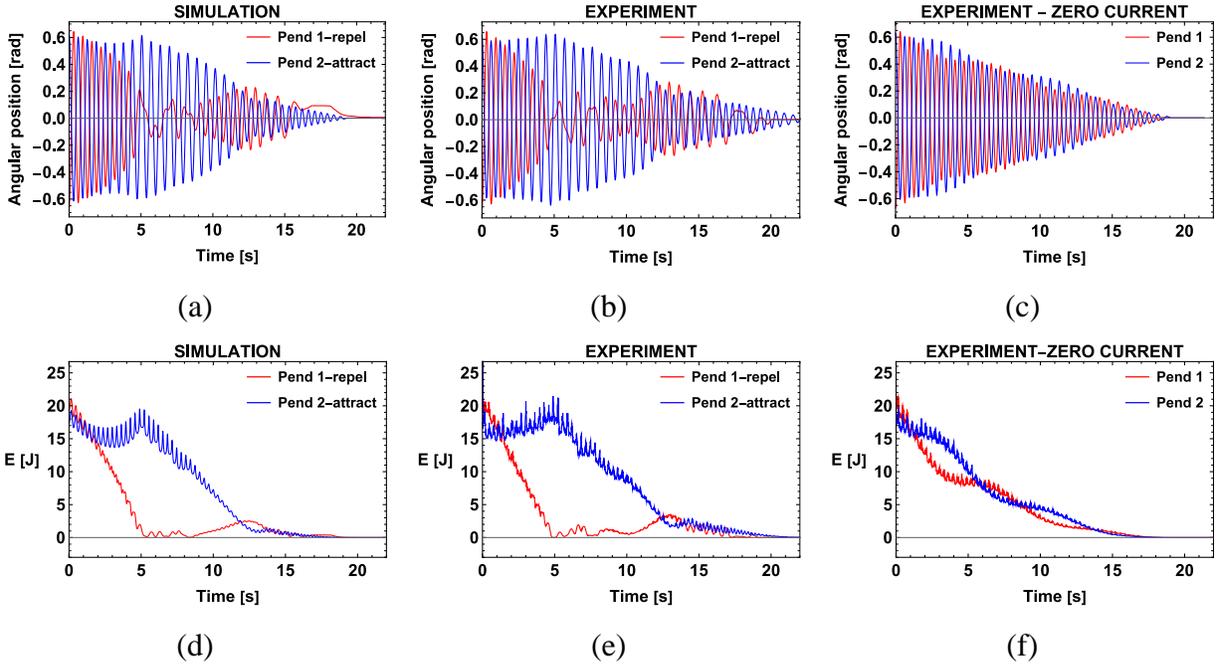

(a) (b) (c)

(d) (e) (f)

Figure 10: Comparison of results of numerical simulations with experimental records. Time histories of the angular displacements (a)-(c), and the corresponding energy distributions versus time (d)-(f). In cases of (a)-(b) and (d)-(e), the input currents are $i_1(t) = 0.001 + 0.4\sin^2(\pi t/11.236)$ and $i_2 = -i_1$, while (c) and (f) illustrate the case of free oscillations under zero input currents $i_1 = i_2 = 0$. Initial conditions are listed in Appendix A and experimental records (e)-(f) were processed with 60-points moving average.

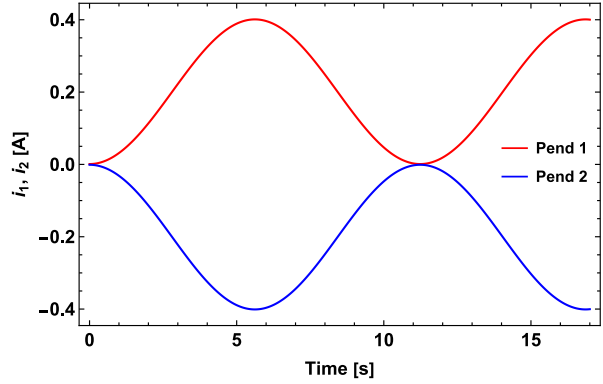

(a)

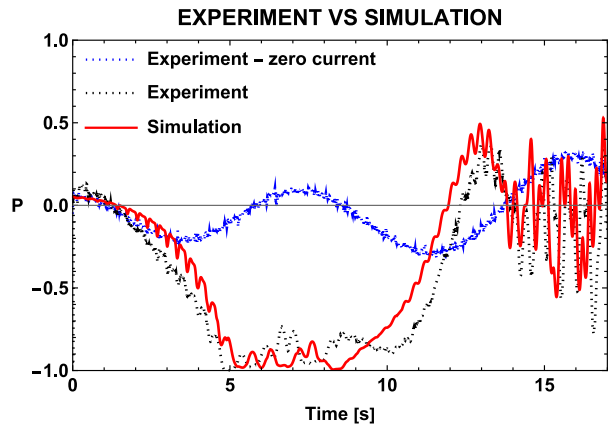

(b)

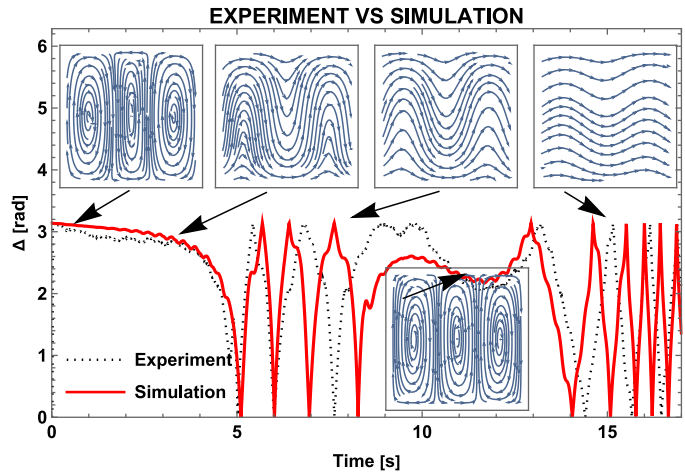

(c)

Figure 11: The effect of a magnetic field in comparison of experimental records with numerical simulations based on the original system (6)-(7) and formulas (9) and (13) obtained for results shown in Fig. 10. Time history of input currents (a), energy distribution $P$ (b) and phase shift $\Delta$ (c); where the meaning of streamlines in square fragments are detailed by Fig. 6 and shown here to explain transitions between qualitatively different temporal intervals of the phase shift, $\Delta$, and the energy distribution, $P$, through the topological changes of the phase portrait in the $P\Delta$– plane.

The presented results for an open-loop controlled system with two input currents clearly exhibit a sufficient correlation between the simulation and experimental results.

## 6. Further system reduction and remarks on the feedback control strategy

The goal of this section is to further reduce the averaged system in order to establish a direct relationship between the input currents and the trend in energy exchange between the pendulums. Then a preliminary qualitative analysis of such relationship will result in a feedback control strategy described in Section 7. Analysis of the right-hand side of system (17) shows that the input currents affect the energy distribution $P$ indirectly through the phase shift $\Delta$. The corresponding explicit mechanism can be revealed by eliminating the phase shift from the system, which is hardly possible to conduct in an exact analytical way due to the strong nonlinearity of the system. Below, we eliminate the phase by linearizing the system separately near both the antiphase mode, $\Delta = \pi$, and inphase mode, $\Delta = 0$. As result, we obtain a second-order linear differential equation for the function $P(t)$ for each of the two cases. The corresponding initial conditions for the derivative $dP/dt$ are given by the second equation in (17) once the initial value $P(0)$ is assigned. Analyzing the derived equations yields to a conclusion regarding analytical dependencies for the feedback control.

### 6.1 *Linearization near the antiphase mode*

Let us assume that the total energy, estimated by $E(t)$, decays slowly in the time scale of energy exchange between the pendulums. Then linearizing the second and third equations of system (17) with respect to the phase near $\Delta = \pi$ and $P$, assuming $E$ fixed, and eliminating the phase gives

$$\frac{d^2P}{dt^2} - 2\Omega\left(2\alpha + \frac{\zeta_1 + \zeta_2}{\pi\sqrt{E}}\right)\frac{dP}{dt}$$
$$+ \left\{\Omega^2\left[\beta^2 + 4\alpha\left(\alpha + \frac{\zeta_1 + \zeta_2}{\pi\sqrt{E}}\right)\right] + \frac{1}{8}\beta E + \frac{a\beta}{Jb^2\Omega^2}e^{-\frac{E}{2b\Omega^2}}\left[(b\Omega^2 + E)I_1 - EI_0\right](i_1 + i_2)\right\}P \quad (20)$$
$$= 8\alpha\Omega^2\frac{\zeta_1 - \zeta_2}{\pi\sqrt{E}} - \frac{a\beta}{Jb}e^{-\frac{E}{2b\Omega^2}}(I_0 - I_1)(i_1 - i_2),$$
$$I_n = I_n\left(\frac{E}{2b\Omega^2}\right) \quad (n=1,2),$$

where $I_n = I_n(z)$ is the modified Bessel function of the first kind, and $i_i = i_i(t)$ ($i$=1,2) are input currents.

The negative coefficient of the effective damping term in Eq. (20) confirms the remark regarding local instability of the spiral stationary points in the cross-sections of the system (17) phase space (recall the discussion of Fig. 6). Also, the difference of currents on the right-hand side of equation (20) reveals that the currents of opposite signs would have a stronger effect on the energy exchange flow.

## 6.2 Linearization near the inphase mode

Following the procedure of the previous subsection with the linearization with respect to the phase near zero gives

$$\frac{d^2P}{dt^2} + 2\Omega\left(2\alpha - \frac{\zeta_1 + \zeta_2}{\pi\sqrt{E}}\right)\frac{dP}{dt}$$
$$+\left\{\Omega^2\left[\beta^2 + 4\alpha\left(\alpha - \frac{\zeta_1 + \zeta_2}{\pi\sqrt{E}}\right)\right] - \frac{1}{8}\beta E + \frac{a\beta}{Jb^2\Omega^2}e^{-\frac{E}{2b\Omega^2}}\left[(b\Omega^2 + E)I_1 - EI_0\right](i_1 + i_2)\right\}P \quad (21)$$
$$= -8\alpha\Omega^2\frac{\zeta_1 - \zeta_2}{\pi\sqrt{E}} + \frac{a\beta}{Jb}e^{-\frac{E}{2b\Omega^2}}(I_0 - I_1)(i_1 - i_2).$$

It is seen that the coefficient of the effective damping term can be just temporally positive until the total energy $E$ is large enough, $\sqrt{E} > (\zeta_1 + \zeta_2)/(2\nu\pi)$.

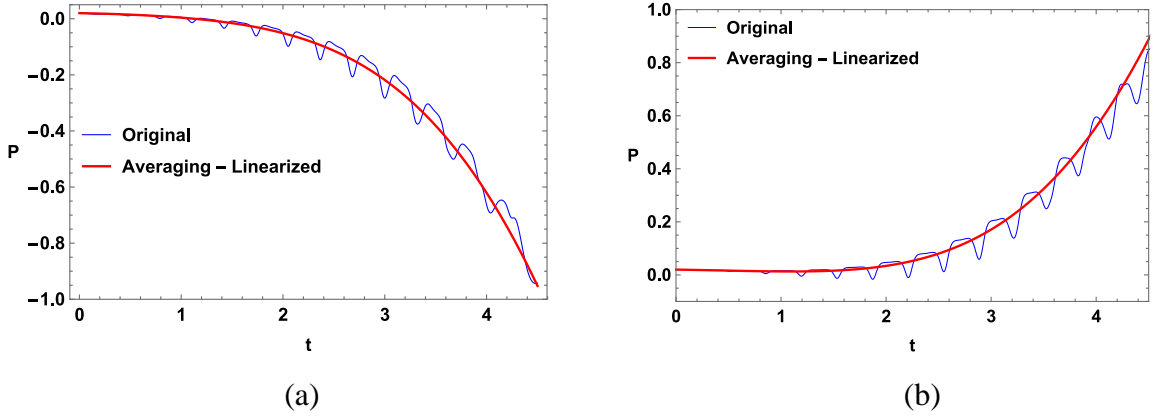

Figure 12: Validation of the averaging and linearization procedures for input currents shown in Fig.11a and two different types of the initial conditions closed to: (a) – antiphase mode, and (b) – inphase mode; see Appendix A for details.

Although linearized Eqs. (20) and (21) were obtained under the assumption $|P| \ll 1$, the results of simulations based on these equations in comparison with numerical solutions of the original system (17) (Figs. 12a and b) reveal that the linearized version remains reasonably applicable even in a wider interval of the energy distribution, $P$. Although the presence of singularities of system (17) at points $P = \pm 1$ invalidates the above linearization in close neighborhoods of these boundary points, a practical significance of this fact is quite low because the goal of control will rather be achieved as soon as the system has reached the neighborhood of one of the singular points for a first time. In case a longer time interval becomes of interest, then a more complicated system (17) still can be used. Recall that Eqs. (9) and (13) are used to process solutions of the original system (6)-(7) in order to switch from the pendulum angles and angular velocities to the energy distribution $P$.

## 7. Feedback control

Analyzing the right-hand sides of the complete averaged system (17), its linearized versions with respect to the parameter of energy distribution (20) and (21), as well as comments in Section 6 leads to the assumption that an effective control strategy for one-directional energy flow between oscillating pendulums can be provided while the coils placed under them are powered by the current signals of opposite signs as functions of the phase shift

$$i_1(t) = -i_0 \cos\Delta, \quad i_2(t) = i_0 \cos\Delta, \tag{22}$$

where $i_0 = const$.

As follows from the third equation in (13), $\cos\Delta$ can be determined from the feedback signals through the elements of the energy matrix (9) as

$$\cos\Delta = Q(t) = \frac{E_{12}(t)}{\sqrt{E_{11}(t)E_{22}(t)}}. \tag{23}$$

Let us emphasize that, in the leading order of averaging, quantity (23) does not depend upon the fast phase of carrying oscillations. The results of simulations are illustrated by Fig. 13, where the current level is $i_0 = 0.35$ in all cases except for (d), where the current was taken $i_0 = 0.5$. The pendulums were equipped with a pair of small magnets and released from different initial positions corresponding to different oscillation modes, i.e., antiphase, inphase, rotating. The values of these initial positions were selected so that for all the shown cases the initial total energy $E$ of the system was the same and equaled 30 J.

It is seen from the left column of Fig. 13 that, under different initial energy distributions and phase shifts between the pendulums, the energy is transferred from the first to the second pendulum, $P \sim -1$. The right column of Fig. 13 confirms that the energy remains with the second pendulum whenever it is initially there. Although there is some fluctuation in the case of the initial 'rotating' phase (d), it can be reduced by increasing the currents amplitude $i_0$. Also, all the fragments of Fig. 13 illustrate a sufficient match between the results of simulations based on the original system (6)-(7) and its averaged version (17).

Recall that the coherency index (23) characterizes the phase shift $\Delta$ between two pendulums as $Q = \cos\Delta$, and therefore is independent on the excitation level. As a result, the strength of feedback (22) is maintained despite the gradual energy dissipation. As follows from the results of numerical simulations (see Fig. 13) such a specific has no undesirable effects on the final phase of dynamics. In reality, different structural imperfections and unaccounted physical factors make the system increasingly sensitive to external inputs as the total energy drops to near equilibrium levels. In order to avoid the related overshoots of control inputs in the neighborhoods of equilibrium states during experiments, we adapt the feedback (22) by introducing the energy dependent decay as

$$i_1(t) = -i_0\{1 - \exp[-\eta E(t)]\}\cos\Delta,$$
$$i_2(t) = i_0\{1 - \exp[-\eta E(t)]\}\cos\Delta, \tag{24}$$

where $E(t)$ is defined in (13), and $\eta$ is a numerical factor, which is selected in an empirical way.

The results of experiments with feedback control (24) were performed for the same system parameters as in numerical simulations and are represented by Fig. 14. They confirm the effects predicted by numerical simulations for three different cases of initial conditions. In particular, Fig. 14a, c, and e show a clear irreversible trend of the energy partitioning index to $P \sim -1$, which is a one-way energy flow to the second pendulum. Note that the intervals of increasing $P$ by the end of time intervals are due to vanishing control currents dictated by the decay factors in (24). Recall that the index $P$ is a relative characteristic. In absolute values, the reverse energy influx is practically ignorable due to very low total energy levels $E$ by the end of time intervals. Besides, the drop of currents is due to the decay of $E$ as follows from the form of feedback (24). Finally, the trajectories on configuration planes (see Fig. 14b, d, and f) confirm the obvious trend to the local mode with $\phi_1 \sim 0$.

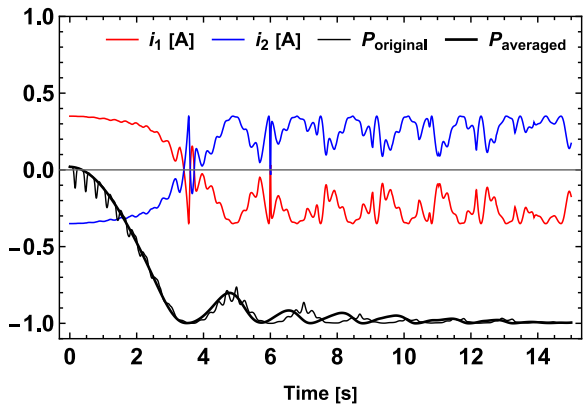

(a)

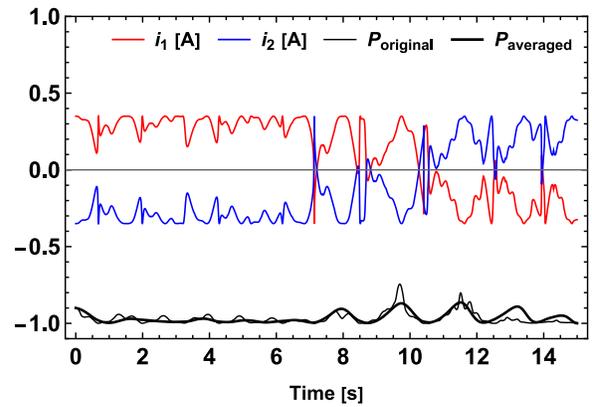

(b)

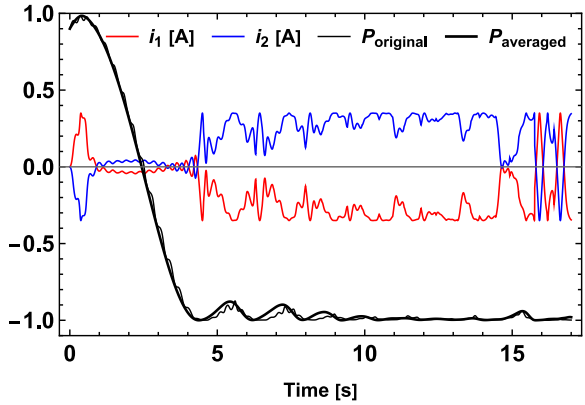

(c)

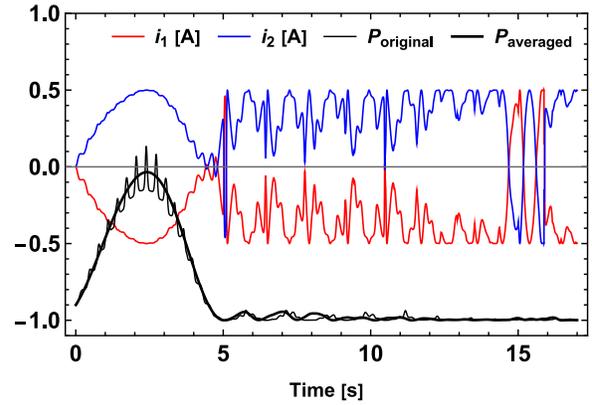

(d)

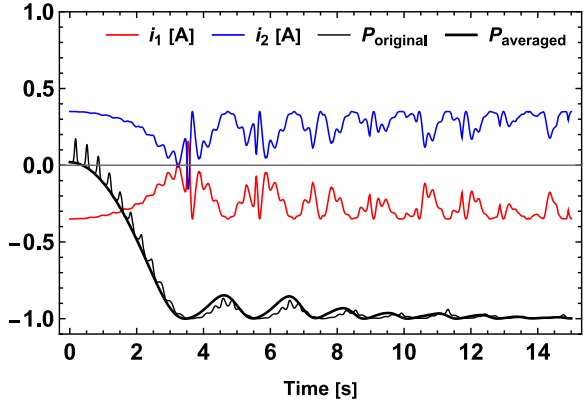
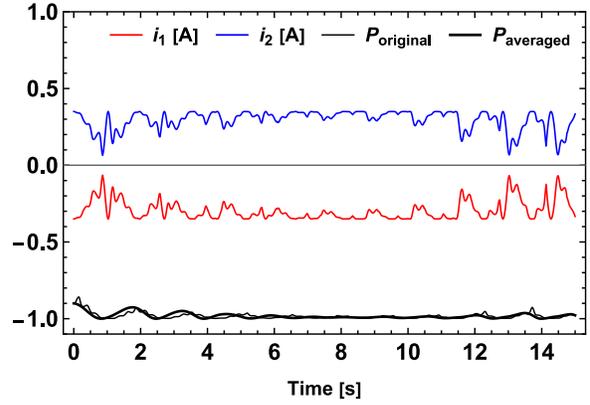

(e)                      (f)

Figure 13: Results of simulations computed for input currents corresponding to the feedback law given by (22) in cases of antiphase mode (a)-(b), 'rotating' mode (c)-(d) and inphase mode (e)-(f). Initial conditions are listed in Appendix A.

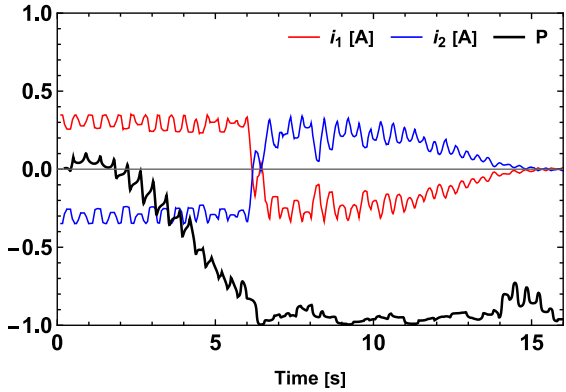
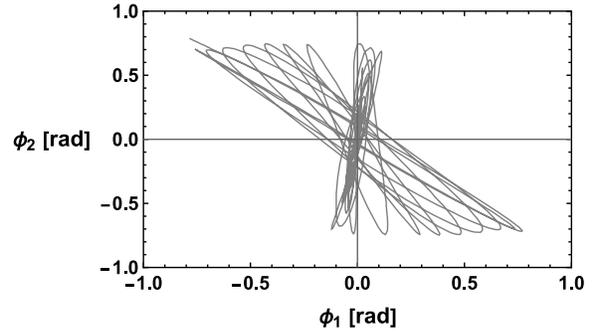

(a)                      (b)

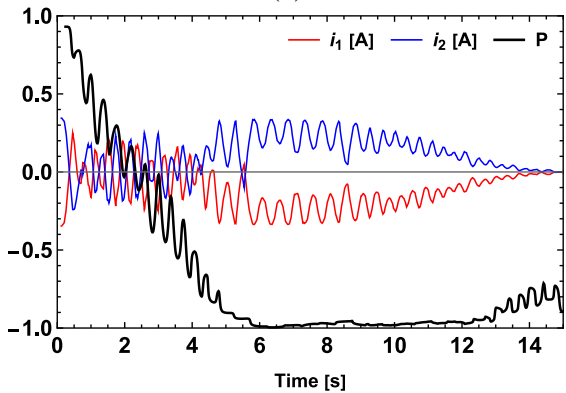
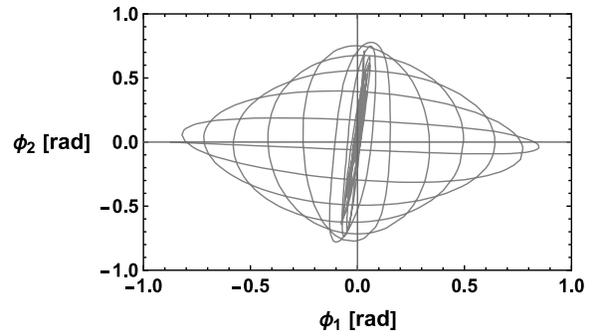

(c)                      (d)

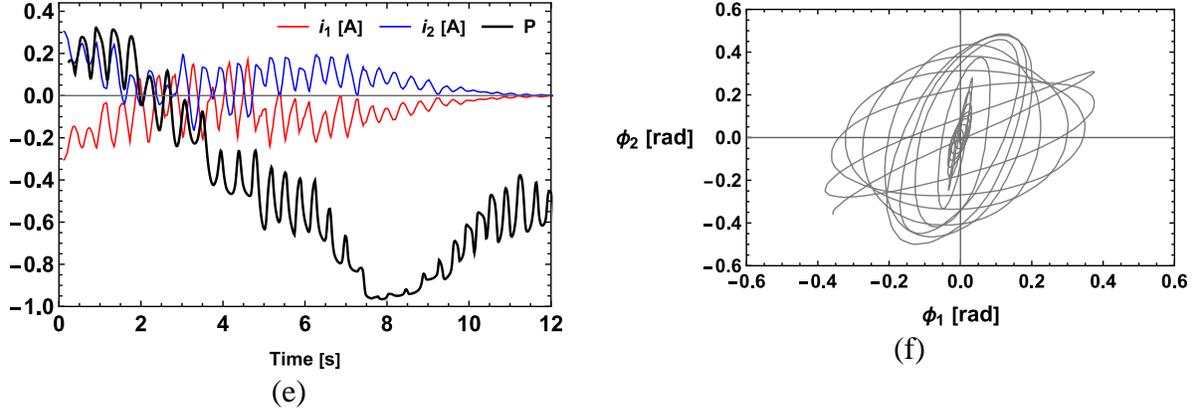

Figure 14: Results of the experiment performed under input currents corresponding to the feedback law given by (24) in cases of antiphase mode (a)-(b), 'rotating' mode (c)-(d) and inphase mode (e)-(f). Initial conditions are listed in Appendix A.

Sample recordings of motion of the controlled system for the case of antiphase mode, inphase mode and rotating mode can be found as supplementary data available online.

## 8. Conclusions

A methodology for guiding the energy flows between two weakly coupled oscillators under 1:1 resonance condition is proposed in the present work. The results of theoretical analyses are validated through a series of experiments showing acceptable agreement between the experimental data and the results of numerical simulations. The corresponding control strategies are revealed from the analysis of coupled compound pendulums in terms of the specific descriptive variables, such as the total excitation level, energy distribution, and the coherency index associated with the phase shift between the pendulums. At the present stage of study, the system is perturbed by an influx of energy through the initial conditions. Then the system remains free of any external loading except for the actuator inputs. The control strategies are based on the physical observation that:

- In the case of antiphase oscillation, the energy is moving *from* the oscillator, which is in the repelling magnetic field, to the oscillator subjected to the attracting field.
- In contrast, during the inphase oscillations, the energy moves *towards* the oscillator under the repelling field.

Both open-loop and closed-loop control strategies are introduced.

- The open-loop control requires details of the initial conditions and the total energy dissipation rate to be a priory known.
- The closed-loop control block requires only the information about the pendulums phase shift, which is estimated from the observed state variables through the coherency index.

Therefore, a closed-loop control method using the feedback may be extended on the case, when the system is subjected to external impacts and other disturbances. The advantage of suggested control strategies is that operational time rates of actuators are dictated by the speed of beating, which is relatively slow as compared to the carrying oscillations. Although the present study deals with a specific physical model of coupled compound pendulums, the methodology is not restrained by the structural specifics of oscillators. The pendulums may have different inertia and geometrical parameters while still maintaining fundamental frequencies in a close to 1:1 proportion. In order to satisfy this condition during the process, the control inputs, dissipative effects, coupling, and nonlinearity should be small compared to the oscillator restoring forces/torques. Structural changes in spatio-temporal scales would require readjustments in the strength of currents and size of magnets. Finally, the present case of coupled identical oscillators may serve as a basis for further extensions of the methodology on pendulum or mass-spring chain models.

## Acknowledgments

The basic construction of the experimental setup presented in the work was supported by the Polish National Science Centre under the grant OPUS 14 No. 2017/27/B/ST8/01330. Moreover, the work of the second author (i.e., experimental measurements and experimental setup modifications, computational data processing and numerical calculations as well as records of experiments) was supported by the National Science Center, Poland under the grant PRELUDIUM 20 No. 2021/41/N/ST8/01019. For the purpose of Open Access, the authors has applied a CC-BY public copyright licence to any Author Accepted Manuscript (AAM) version arising from this submission. This is an AAM of an article published by Elsevier in Mechanism and Machine Theory on 19 July 2022, available at: https://doi.org/10.1016/j.mechmachtheory.2022.105019

## Appendix A

This appendix contains the values of the initial conditions for which the results presented in the figures were obtained.

Fig. 5   Initial conditions: $\phi_1(0) = 0.568$ rad; $\phi_2(0) = -0.557$ rad; $v_1(0) = 0$ rad/s; $v_2(0) = -5.166 \cdot 10^{-3}$ rad/s.

Fig. 7   Initial conditions: $\phi_1(0) = -0.545$ rad; $\phi_2(0) = 0.580$ rad.

Fig. 8   Initial conditions: $\phi_1(0) = -0.697$ rad; $\phi_2(0) = 0.404$ rad.

Fig. 10  Initial conditions: $\phi_1(0) = -0.660$ rad; $\phi_2(0) = 0.627$ rad.

Fig. 12  Initial conditions: $E(0) = 30.0$ J; $P(0) = 0.02$; antiphase mode $\Delta(0) = \pi - 0.001$ rad; inphase mode $\Delta(0) = -0.001$ rad.

Fig. 13  Initial conditions: $E(0) = 30.0$ J; $P(0) = 0.02$; antiphase mode $\Delta(0) = \pi - 0.001$ rad; 'rotating' mode $\Delta(0) = \pi/2 - 0.001$ rad; inphase mode $\Delta(0) = -0.001$ rad.

Fig. 14  Initial conditions: antiphase mode, A, $\phi_1(0) = \pi/4$ rad, $\phi_2(0) = -\pi/4$ rad; 'rotating' mode, A, $\phi_1(0) = 5\pi/18$ rad, $\phi_2(0) = 0$ rad; inphase mode, A, $\phi_1(0) = \phi_2(0) = \pi/9$ rad.